\documentclass[twocolumn,showpacs,preprintnumbers]{revtex4}
\usepackage{amsmath,amsthm}
\usepackage{graphicx}
\usepackage{dcolumn}
\usepackage{bm}

\begin{document}


\title{Constrained Molecular Dynamics Simulations of Atomic Ground-States} 
\author{Sachie Kimura}
\author{Aldo Bonasera}
\affiliation{Laboratorio Nazionale del Sud, INFN,
via Santa Sofia, 62, 95123 Catania, Italy}
\date{\today}

\begin{abstract}
Constrained molecular dynamics(CoMD) model, previously introduced for nuclear 
dynamics, has been extended to the atomic structure and collision calculations. 
Quantum effects corresponding to the Pauli and Heisenberg principle are 
enforced by constraints, following the idea of the Lagrange multiplier method.   
Our calculations for small 
atomic system, H, He, Li, Be, F reproduce the ground-state binding energies 
reasonably, compared with the experimental data.
We discuss also the shell splitting which is expected as a consequence of the 
e-e correlation.  
\end{abstract}


\maketitle

Molecular dynamics approach is a powerful tool to simulate 
nuclear heavy ion collisions and atomic collisions, due to 
its simplicity and ability to take into account the influence of 
correlations and fluctuations. 
However, as it is seen in the case of classical trajectory Monte Carlo 
method~\cite{ap}, 
truly classical atoms, without constraints due to the Heisenberg 
and Pauli principle, are unstable. To describe 
the ground-state properties of the systems with molecular dynamics approach, 
the pseudo potential is often introduced to simulate the effects of the Heisenberg 
uncertainty principle and the Pauli exclusion principle~\cite{kw,cohen0}. 
The method with pseudo potential is known as Fermionic molecular dynamics(FMD) and 
it is applied to the studies of various atomic processes, in which 
a fully quantum mechanical dynamical simulation encounters numerical problems difficult
to overcome. To give some 
actual examples, the atom-ion collisions~\cite{bw}, the atomic ionization by laser 
fields~\cite{wk}, the capture of muons by 
hydrogen atoms~\cite{cohen}, the formation of antiprotonic atom~\cite{cohen2}
and the ionization cross section calculations~\cite{gr},
all these processes have been investigated using molecular dynamics approach.
In particular, Geyer and Rost~\cite{gr} proposed the quasi-classical calculation 
of the ionization cross section using classical propagation scheme.
As the initial state of the target, they use the phase space distributions of the 
bound electrons which are directly obtained from quantum mechanical wave functions
through the Wigner transformation.  There they mention the importance of the 
energy spread of the phase space distributions. 

Meanwhile, a constrained molecular dynamics (CoMD) approach has been proposed to 
treat fermionic properties of the nucleons in nuclei~\cite{pmb}.  
The approach has been successfully applied to study the Equation of State of 
the quark system as well~\cite{tb}.  
In this approach, the Pauli exclusion principle is 
accomplished by restricting the phase space occupancy $\bar{f}_i$ to values less or 
equal to 1~\cite{bgm}. 
The equation of motion with the constraints for each electron is derived on the basis of 
Lagrange multiplier method for constraints. 
The constraint of CoMD approach is thought as an alternative to the  
pseudo potential and can be easily extended to the case of the Heisenberg 
uncertainty principle as well.  
The constraints play the role of a ``dissipative term'' in the 
classical equation of motion and lead the system to its ground-state.
But in CoMD, at variance with FMD approaches, the ``dissipative term'' can increase or decrease
the energy of the system depending on the phase space occupation.

In this brief report, we apply CoMD to atomic systems for the purpose of determining 
their ground-states configurations.
Particularly, we discuss some properties of ground-states atoms, i.e., binding 
energies (the total electronic energies) and radial positions of the bound electrons. 
We discuss also the energy spread (variance of the binding energy) and radial variance
of each bound electron as well.  
In our approach, we prepare the ensemble of initial configurations.
The binding energies and radial positions of the electrons in the systems are 
calculated as averaged values over the ensembles.
Our approach is sufficient to obtain stable atomic 
ground-states providing their atomic energies fairly accurately.   
Using the obtained ensembles of initial states which occupy different points 
in the phase space, molecular dynamics simulation with constraints for the atomic 
collision has been performed and applied to nuclear fusion enhancement factor calculations 
of D+$d$ reaction for astrophysical interests~\cite{kb}. 

We describe the essence of CoMD briefly. 
In classical molecular dynamics(CMD) one solves the Hamilton equations, i.e.,: 
\begin{align}
  \label{eq:rt}  
  \frac{d {\bf r}_i}{dt}= \frac{{\bf p}_i c^2}{{\mathcal E}_i};\hspace{0.8cm}
  \frac{d {\bf p}_i}{dt}= -\nabla_{{\bf r}} U({\bf r}_i), 
\end{align}
where we use relativistic kinematics: 
${ \mathcal E}_i=\sqrt{{\bf p}_i^2c^2+m_i^2c^4}$, $m_i$ and 
\begin{equation}
  \label{eq:coulomb}
  U({\bf r}_i)=\sum^N_{j(\neq i)=0} \frac{q_j q_i }{|{\bf r}_i-{\bf r}_j|}  
\end{equation}
are the energy, the mass and the potential of the $i$-th electron, respectively.
Here, $q_{i}$ is the charge of the electron $i$ and $q_j$ is the charge of 
electron(for $1\le j\le N$) or nucleus(for $j=0$).  
The total Hamiltonian is written down as 
$H({\bf r};{\bf p})=\sum_i^N ({ \mathcal E}_i+U({\bf r})-mc^2)$.

The shortcoming of the approach is the lack of 
the Pauli exclusion principle. CoMD gives a possibility to overcome the shortcoming. 
To take the feature of the Pauli blocking into account,  
we use the Lagrange multiplier method for constraints.
Our constraints which correspond to the Pauli blocking is 
$\bar f_i\le 1$ in terms of the occupation probability and 
can be directly related to the distance of two particles, i.e.,
$r_{ij} p_{ij}$, in the phase space.
Here $r_{ij}=|{\bf r}_i-{\bf r}_j|$ and 
$p_{ij}=|{\bf p}_i-{\bf p}_j|$. 
The relation $\bar f_i \le 1$ is fulfilled, if $r_{ij} p_{ij}\ge
\xi_P\hbar\delta_{s_i,s_j}$, where $\xi_P=2\pi(3/4\pi)^{2/3}2^{1/3}$, $i,j$ refer only to electrons and 
$s_i,S_j(=\pm 1/2)$ are their spin projection. 
We can easily extend the approach to 
the Heisenberg principle where the constraint is expressed as
${r}_{ij} {p}_{ij}\ge\xi_H\hbar$, where $\xi_H=1$,
$i$ and $j$ refer to the electrons and the nucleus.
Using these constraints, the Lagrangian of the system can be written as 
\begin{align}
   {\mathcal L}=\sum_i^N {\bf p}_i \cdot \dot{\bf r}_i &-H({\bf r}; {\bf p})\nonumber 
   + \sum_{i,j(i)}\lambda^H_i \left( \frac{{r}_{ij} {p}_{ij}}{\xi_H \hbar}-1 \right) \\
   &+ \sum_{i,j(i)}\lambda^P_i \left( \frac{{r}_{ij} {p}_{ij} \delta_{s_i,s_j}}{\xi_P \hbar}-1 \right),  \label{eq:lagc}
\end{align}
where $\lambda^P_i$ and $\lambda^H_i$ are Lagrange multipliers for Pauli and Heisenberg principle 
respectively.
The variational calculus leads to:
\begin{align}
  \label{eq:rt2}  
  \frac{d{\bf r}_i}{dt} &= \frac{{\bf p}_i c^2}{{\mathcal E}_i} 
  + \frac{1}{\hbar}\sum_{j(i)}\left(\frac{\lambda_i^H}{\xi_H}
    +\frac{\lambda_i^P}{\xi_P}\delta_{s_i,s_j}\right) 
  {r}_{ij}\frac{\partial {p}_{ij}}{\partial {\bf p}_i }, \\
  \label{eq:pt2} 
  \frac{d {\bf p}_i}{dt}&= -\nabla_{{\bf r}} U({\bf r}_i)
  - \frac{1}{\hbar}\sum_{j(i)}\left(\frac{\lambda_i^H}{\xi_H}
    +\frac{\lambda_i^P}{\xi_P}\delta_{s_i,s_j}\right) 
  {p}_{ij}\frac{\partial {r}_{ij}}{\partial {\bf r}_i }. 
\end{align}
From physical considerations we expect that the Pauli principle is 
stronger than the Heisenberg principle 
for the two closest identical electrons in the phase space, i.e., the 
particles $i$ and $j(i)$ for which $r_{ij}p_{ij}$ is smallest. While 
the Heisenberg principle must be enforced especially among the electrons and 
the nucleus. Thus we restrict the summations in eqs.~\eqref{eq:lagc}-\eqref{eq:pt2}        
to those particles only.

In order to obtain the atomic ground-state configuration, 
we perform the time integration of the eqs.~\eqref{eq:rt2} and \eqref{eq:pt2}.
The values of $\lambda_i^H$ and $\lambda_i^P$ are determined depending on the 
magnitude of ${r}_{ij}{p}_{ij}$. If 
${r}_{ij}{p}_{ij}$ is (smaller)larger than $\xi_{H(P)} \hbar$, 
$\lambda$ has positive(negative) sign, changing the phase space 
occupancy of the system.  
The constraints work as the ``dissipative term'' in the case of the pseudo potential 
approach and lead automatically to the minimum energy, i.e., the ground-state of the system.
The difference being that in the case of the model with the pseudo potential, a dissipative 
term decreases the total energy.
In our case the total energy decreases or increases depending on the phase space occupancy.  

The electron configurations at the beginning of the time integration are prepared 
in the following way.
In the case of even number of bound electrons, we 
locate a pair of them at the opposite points respect to the nucleus in the phase space.
In this way, the center of mass of the electrons coincides with the position 
of the nucleus, i.e., total momentum of the electrons is zero.  
We do the same procedure for the odd-number electrons atom, 
excluding an electron which is the outermost. 
Thus, at the beginning of the time integration we have an ensemble 
of electron configurations which occupy different points in the phase space microscopically. 
The integration of the eqs.~\eqref{eq:rt2} and \eqref{eq:pt2} is performed using Hermite integration 
scheme which is efficient and enables integration with high precision. The scheme adopts variable 
and individual time-steps for each electron~\cite{ma}.
Considering the nucleus rest frame, 
The binding energy of the atom is determined by 
\begin{equation}
  \label{eq:ave}
  B.E.=\frac{1}{nev}\sum^{nev} H({\bf r}; {\bf p}),
\end{equation}
where $nev$ is the number of the events in the ensemble. 
The radial coordinates of the electron $i$ ($R_i$) is given by
\begin{equation}
  \label{eq:rc}
  R_i=\frac{1}{nev}\sum^{nev}\sqrt{|{\bf r}_i|^2}.
\end{equation}
The binding energies and the radial coordinates of the electron $i$
of the atoms are calculated as an averaged value over the ensemble of events.

\begin{figure}[htbp]
  \centering
  \includegraphics[width=7.5cm,clip]{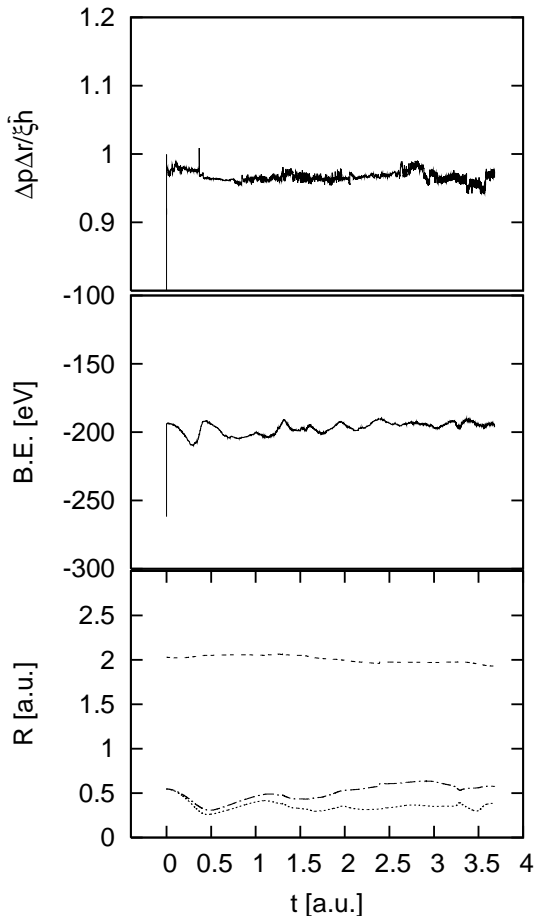}
  \caption{The convergence to the ground state of the lithium atom. The values are averaged over 18 events.}
  \label{fig:average}
\end{figure}

\begin{figure}[htbp]
  \centering
  \includegraphics[width=7.5cm,clip]{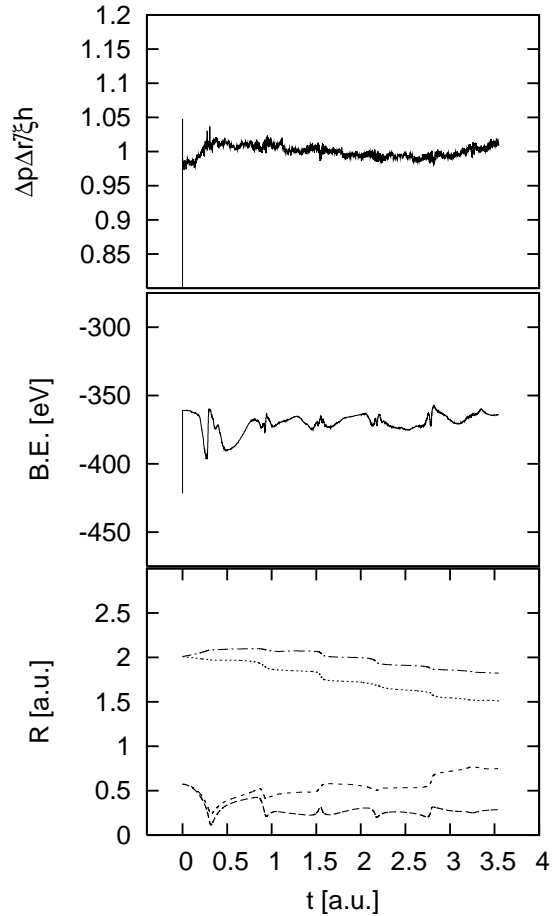}
  \caption{The convergence to the ground state of the beryllium atom. The values are averaged over 20 events.}
  \label{fig:average_be}
\end{figure}

We have applied the model to hydrogen, helium, lithium, beryllium, fluorine atoms.  
Fig.~\ref{fig:average} and Fig.~\ref{fig:average_be} show that the systems converge to 
their ground-state in the illustrative cases of lithium and beryllium atoms, respectively.   
The top panels show the time development of the average of 
${r}_{ij} {p}_{ij}/\xi \hbar$ over all pair of particles and 
over events in the ensemble, we write it as $\Delta {\bf r} \Delta {\bf p}/\xi \hbar$. 
The middle panels and the bottom ones show the average of binding energy and the 
radial coordinates of the electrons, respectively, over events. 
In the bottom panels each line corresponds to the radius of each bound electron.  
Due to the constraints these values oscillate as a function of time and converge 
after some time. 
We determine the binding energy and the radius by taking the average over 
not only events but also over time.    

We summarize our results of the ground-state energies and the radial coordinates of 
the bound electrons  
for small atomic systems in Table~\ref{tab:atoms} and~\ref{tab:atomsr}, respectively.   
For the purpose of utilizing the atomic configuration to collision calculations,
the comparison between our results and the ones from quantum mechanical 
Hartree-Fock(HF) is suggestive. 
In Table~\ref{tab:atoms} together with the ground-state energies from our method,  
results from the FMD~\cite{cohen1}, HF~\cite{fischer} methods and experimental 
values~\cite{ben} are shown. 
Since we determine the binding energies as an average of many events, our results have 
a variance $\sqrt{(\Delta B.E.)^2}$, also included in the tables. 
We obtain ground-state binding energies in good agreement with experimental data 
within variances.
In Table~\ref{tab:atomsr} we compare our results of the radial coordinates of 
the electrons($R_i$) for each atoms with FMD and HF methods. The results 
from FMD which are obtained using different parameter sets are shown in two 
columns(FMD~0~\cite{cohen0} and FMD~1~\cite{cohen1}). 
The column FMD~1 is with the optimized parameter sets.       
Note that our method gives smaller values as $R_i$ than those from HF method.
However one should notice that while comparison of our approach to FMD 
is direct since they are all semi-classical molecular dynamics, thus the 
definition given in eqs.~\eqref{eq:ave},~\eqref{eq:rc} are exactly the same.  
On the other hand in the HF approach there are no ``events", the calculated values 
of HF are obtained through a smooth probability distribution, which is not the 
case for our model where we have $\delta$-functions. 
Furthermore symmetries imposed to the system will be preserved in HF, indeed they 
might be destroyed in CoMD because of correlations. 
We give an example of the Be case; initially we distributed the electrons pairwise, one 
opposite to the other respect to the nucleus both in the coordinate 
and momentum space, as stated above. This initial symmetry is broken in the simulation 
because of e-e correlation, nevertheless one can recognize two major shells within the 
variance $\Delta R_i$ in the results.
Certainly even in our calculations we can impose such a symmetry. 
In Tables~\ref{tab:atoms} and~\ref{tab:atomsr} the numbers in parenthesis show
in the case where we impose that 
electrons move pairwise opposite locations in the phase space, i.e., they are forced 
to keep the initial symmetries. Such a calculation gives a very similar $B.E.$ as 
in the absence of forced symmetry, but now two major shells are clearly identified.  
From the comparison of these two cases it is obvious 
the splitting of the shells when relaxing the initial symmetry.

\begin{table*}
\caption{\label{tab:atoms}Summary of the binding energies (in eV) of calculated systems.
The binding energies with CoMD including error bars, with optimized FMD(FMD~1)~\cite{cohen1}, 
with Hartree-Fock method~\cite{fischer} and experimental values~\cite{ben} are shown 
for each atom. }
\begin{ruledtabular}
\catcode`?=\active \def?{\phantom{0}}
\renewcommand{\arraystretch}{1.2}
\begin{tabular}{llllll}
   & ??????CoMD     & ??FMD~1      & ????HF     & experimental \\[-1pt] 
\hline
H  & ??$-$13.3 $\pm$?1.8 & ??$-$13.61 & ??$-$13.61 &  ??$-$13.6  \\[-1pt] 
He & ??$-$79.1 $\pm$?6.9 & ??$-$77.57 & ??$-$77.87 &  ??$-$79.1  \\[-1pt] 
Li & ?$-$196.8 $\pm$?15.& ?$-$197.61 & ?$-$202.26 & ?$-$203.6  \\ [-1pt] 
Be & ?$-$371.2 $\pm$?27. ($-$385.3 $\pm$?29.)\footnote{with forced symmetry}& ?$-$395.63 & ?$-$396.56 & ?$-$399.4  \\
F  &  $-$2560.3 $\pm$192.& $-$2408.3  & $-$2705.1 & $-$2717.4  \\ [-1pt] 
\end{tabular}
\end{ruledtabular}
\caption{\label{tab:atomsr}Summary of the radial coordinates of bound electrons(rc) 
(in atomic unit) of calculated systems. The rc with CoMD, with FMD(FMD~0~\cite{cohen0} and  
optimized FMD~1~\cite{cohen1}) and with Hartree-Fock method~\cite{fischer} 
are shown.}
\begin{ruledtabular}
\catcode`?=\active \def?{\phantom{0}}
\renewcommand{\arraystretch}{1.2}
\begin{tabular}{lllll}
 & CoMD & FMD~0 ???FMD~1 & HF  & ???????\\ 
\hline
H  & 1.0 $\pm$ 0.2 & 1.0??? ??? 1.0??? & 1.0???(1s) & \\ [2pt]
He & 0.55 $\pm$ 0.09 & 0.5714 ??? 0.6139 & 0.9273(1s) & \\ [-2pt]
   & 0.60 $\pm$ 0.1 & 0.5714 ??? 0.6139 &        & \\ [2pt]
Li & 0.34 $\pm$ 0.1 & 0.3506 ??? 0.3818 & 0.5731(1s) & \\ [-2pt]  
   & 0.5 $\pm$ 0.2 & 0.3673 ??? 0.3885 & 3.8737(2s) & \\ [-2pt]
   & 2.0 $\pm$ 0.6 & 1.4419 ??? 4.0604 &            & \\ [2pt]
Be & 0.27 $\pm$ 0.13? (0.36 $\pm$ 0.17)\footnote{same as in Table~\ref{tab:atoms},with forced symmetry} & 0.2565 ??? 0.2687 & 0.4150(1s) & \\ [-2pt] 
   & 0.47 $\pm$ 0.20? (0.36 $\pm$ 0.17) & 0.2565 ??? 0.2687 & 2.6494(2s) & \\ [-2pt]
   & 1.63 $\pm$ 0.45? (1.60 $\pm$ 0.34) & 0.9458 ??? 2.9738 &            & \\ [-2pt]
   & 1.83 $\pm$ 0.47? (1.60 $\pm$ 0.34) & 0.9458 ??? 2.9738 &        & \\ [2pt]
F  & 0.15 $\pm$ 0.09  & ?????? ??? 0.2625 & 0.1757(1s) & \\ [-2pt]
   & 0.30 $\pm$ 0.09  & ?????? ??? 0.2624 & 1.0011(2s) & \\ [-2pt]
   & 0.41 $\pm$ 0.05  & ?????? ??? 0.2622 & 1.0848(2p) & \\ [-2pt]
   & 0.48 $\pm$ 0.05 & ?????? ??? 0.2622 &        & \\ [-2pt]
   & 0.55 $\pm$ 0.07 & ?????? ??? 0.2668 &        & \\ [-2pt]
   & 0.61 $\pm$ 0.07 & ?????? ??? 0.2671 &        & \\ [-2pt]
   & 0.69 $\pm$ 0.07 & ?????? ??? 0.6931 &        & \\ [-2pt]
   & 0.75 $\pm$ 0.07 & ?????? ??? 0.6952 &        & \\ [-2pt]
   & 0.87 $\pm$ 0.1 & ?????? ??? 0.7818 &        & 
\end{tabular}
\end{ruledtabular}
\end{table*}

We have presented results of Constrained molecular dynamics approach 
to describe the atomic ground-states configurations.
We calculated the binding energies and the radial coordinates of the electrons in atoms. 
The total electronic energies for the ground-state atoms are given rather accurately,
At last we stress that the intent of the CoMD simulation of the atomic systems is to applying 
it to the collision calculations and determining the Equation of State of matter at very 
low temperatures where quantum effects play a decisive role.

\smallskip
We thank Prof. J.S. Cohen for providing us his data.


\end{document}